\documentstyle[12pt]{article}

 \newcommand{\be}{\begin{equation}}
 \newcommand{\ee}{\end{equation}}
 \newcommand{\ba}{\begin{eqnarray}}
 \newcommand{\ea}{\end{eqnarray}}
 
 \newcommand{\del}{\partial}

\def\lt{\tilde \lambda}

\def\infinity{\infty}

\newcommand{\lef}{\left}

\newcommand{\ri}{\right}

\newcommand{\cl}{{\cal L}}

\newcommand{\fr}{\frac}

\begin{document}

\begin{titlepage}

\topmargin -15mm

\vskip 10mm
\vskip 25mm

\centerline{ \LARGE\bf Different Approaches for Bosonization }
\vskip 2mm
\centerline{ \LARGE\bf  in Higher Dimensions }

    \vskip 2.0cm

    \centerline{\sc R.Banerjee $^*$ and E.C.Marino }

     \vskip 0.6cm
     
\centerline{\it Instituto de F\'\i sica}
\centerline{\it Universidade Federal do Rio de Janeiro } 
\centerline{\it Cx.P. 68528, Rio de Janeiro, RJ 21945-970, Brasil} 
\vskip 2.0cm

\begin{abstract} 

We describe two distinct approaches for bosonization in higher dimensions;
one is based on a direct comparison of current correlation functions
while the other relies on a Master lagrangean formalism. These are used
to bosonize the Massive Thirring Model in three and four dimensions
in the weak coupling regime but with an arbitrary fermion mass. In both
approaches the explicit bosonized lagrangean and current are derived in
terms of gauge fields. The complete equivalence of the  two bosonization
methods is established. Exact results for the free massive fermion theory
are also obtained. Finally, the two-dimensional theory is revisited and
the possibility of extending this analysis for arbitrary dimensions is
indicated.
                                     
\end{abstract}

\vskip 3cm
$^*$ On leave of absence from S.N.Bose National Centre for Basic
Sciences, Calcutta, India.
\vskip 3mm
Work supported in part by CNPq-Brazilian National Research Council.
     E-Mail addresses: rabin@if.ufrj.br; marino@if.ufrj.br

\end{titlepage}

\hoffset= -10mm

\leftmargin 23mm

\topmargin -8mm
\hsize 153mm
 
\baselineskip 7mm
\setcounter{page}{2}

\section{Introduction}

Bosonization is a very powerful method by which fermionic theories
are mapped into bosonic ones. It was first created and fully developed
in the realm of two-dimensional physics. For some time it was
thought that bosonization was only possible because of the strict
constraint imposed by the dimensionality of space. Further research,
however, revealed a deep structure underlying the process of bosonization.
In particular, it became
clear that a fundamental feature of this process is the fact that the
basic fermions are mapped into the topological excitations of the
associated bosonic theory. Correspondingly, the fermionic current is
mapped into the topological current in the bosonic version and therefore
its conservation is automatically implied. 
More recently, it has been found that this structure
was not restricted to any specific dimension. The idea of
bosonization was then generalized to higher dimensions by different
methods \cite{em1,rb,rb1,bos,rb2,bfo,bm,bm2}.

One of these methods which has been useful in studying abelian
bosonization is the so called Master Lagrangean formalism
\cite{rb,rb1,bos,bm}. This is based
on the coupling of the fermion to a dynamical boson gauge field which is a
vector in three dimensions, a second rank antisymmetric tensor in
four dimensions and so on. Integration over the bosonic degrees of
freedom leads to the original fermionic theory, whereas integration
over the fermion field yields the corresponding bosonized theory.
Another approach which has been exploited even for the nonabelian
bosonization is the hamiltonian constrained formulation \cite{rb2}.
This consists in
embedding techniques which allow to convert the original
fermionic theory into its bosonized form. An outline of a third method
\cite{bm2}
was recently introduced which is based on the direct comparison of
current correlation functions. This will be fully developed and
exploited in the present work. In spite of the variety of bosonization
methods in dimensions higher than two, a direct operator bosonization
of the fermion field which leads to a complete description of the
fermion correlators, as well as the expressions for
the lagrangian and current operators
in the framework of the associated bosonic theory
was only obtained \cite{em1} in the case
of a free massless fermion in 2+1D. On the
other hand, the connection among the different bosonization methods
has not been established. Among other points which could be stressed,
one is the fact that
the bosonization of an interacting theory for an arbitrary fermion mass
was still missing, despite the number of available techniques. Another
one is the lack of any discussion on the
apparent nonrenormalizability of the
fermionic theory in contradistinction to its bosonic counterpart.

In this work, we address some of the important issues raised above.
The method of current correlators is applied to the bosonization of
the Massive Thirring Model (MTM) both in three and four dimensions
in the weak coupling regime but with an arbitrary fermion mass.
Explicit forms for the bosonized current and lagrangian are obtained
in terms of a generalized free vector gauge field. From the general
structure of these expressions one is able to conclude that similar
results hold in arbitrary dimensions. The issue of the nonrenormalizability
of the MTM and its implications for the bosonization are analyzed.
We then reconsider the bosonization of the MTM by means of the Master
Lagrangean approach. Again, the explicit expressions for the bosonized
current and lagrangean both in three and four dimensions are derived
but now these are given, respectively, in terms of
a vector gauge field and of a second rank antisymmetric tensor gauge field.
In both cases they are also generalized free gauge fields in the weak
coupling regime of the MTM. This clearly shows
the difference from the approach
based on the comparison of current correlators where the bosonic field
is always a vector. Nevertheless, we shall explicitly demonstrate the
complete equivalence between the two bosonization schemes. In particular,
the mapping connecting the bosonic gauge fields in the two methods
is obtained. At the end, we revisit the bosonization of two-dimensional
Massless Thirring Model, through the application of both the above mentioned
methods, showing how the well known results are easily reproduced.

\section{Massive Thirring Model in Three Dimensions}

In this section we investigate the bosonization of the MTM by first developing
the recently introduced approach based on the direct comparison of
current correlation functions \cite{bm2} and then by the so called Master
Lagrangean approach \cite{rb,rb1,bm}. We also explicitly show the
equivalence between the two methods.

\subsection{  Current Correlator Approach}
\setcounter{equation}{0}

Our starting point is the current correlators generating functional
in the euclidean space
\be
Z[J] = \int  D\psi D\bar\psi
\exp \lef\{-\int d^3z \lef [ 
\bar\psi (- \not\!\del + m ) \psi - \fr{\lambda^2}{2} j_\mu j_\mu +
i\lambda j_\mu  J_\mu \ri ] \ri \}
\label{z1}
\ee
where $j^\mu = \bar\psi \gamma^\mu \psi$ and we follow
the metric of \cite{cl}.

The current-current interaction can be described in terms of an
auxiliary vector field in the following way
\be
Z[J] = \int  D\psi D\bar\psi D A_\mu
\exp \lef\{-\int d^3z \lef [ 
\bar\psi (- \not\!\del + m + i\lambda \not\!\!\! A ) \psi
- \fr{1}{2} A_\mu A_\mu +
i\lambda j_\mu  J_\mu \ri ] \ri \}
\label{z2}
\ee
In the weak coupling approximation, only the 
two-legs one-loop graph contributes to the fermion determinant
so that 
\be
Z[J] = \int   D A_\mu
\exp \lef\{-\int d^3z \lef \{ 
 \fr{ \lambda^2}{2} ( A_\mu + J_\mu ) \Pi_{\mu\nu} ( A_\nu + J_\nu )
- \fr{1}{2} A_\mu A_\mu 
 \ri \} \ri \}
\label{z3}
\ee
The one-loop
vacuum polarization tensor $\Pi_{\mu\nu}$ has been computed both
by using lattice  \cite{cl} and continuum \cite{djt} regularizations,
giving in momentum space
\be
\Pi_{\mu\nu} (q) = A(q^2) C_{\mu\nu}(q) + B(q^2) P_{\mu\nu}(q)
\label{pi}
\ee
where $C_{\mu\nu}(q) = \epsilon_{\mu\nu\alpha} q_\alpha$ and
$P_{\mu\nu}(q) = q^2 \delta_{\mu\nu}
- q_\mu q_\nu$, while
\be
A(q^2 ) = a_0 + \fr{1}{4\pi} \int_0^1 dt \{ 1 - m [ m^2 + t(1-t) q^2
]^{-1/2} \}
\label{1a}
\ee
and
\be
B(q^2) = 
 \fr{1}{2\pi} \int_0^1 dt t(1-t)  [ m^2 + t(1-t) q^2 ]^{-1/2}
\label{b}
\ee
Note that the parity violating term, as is well known, contains a
regularization dependent finite term  proportional to $a_0$
\cite{cl, djt}.

The following algebraic relations among $C_{\mu\nu}$ and $P_{\mu\nu}$
will prove to be very useful in what follows:
\be
C_{\mu\alpha}C_{\alpha\nu} = - P_{\mu\nu} \ \ ;\ \ 
P_{\mu\alpha}P_{\alpha\nu} = q^2  P_{\mu\nu} \ \ ;\ \  
C_{\mu\alpha}P_{\alpha\nu} =
P_{\mu\alpha}C_{\alpha\nu} = q^2  C_{\mu\nu}
\label{cp}
\ee

After integrating over $A_\mu$ in (\ref{z3}), we obtain
\be
Z[J] = 
\exp \lef\{-\int d^3z  \fr{ \lambda^2}{2} \lef \{ 
   J_\mu  \Pi_{\mu\nu}  J_\nu 
+ \lambda^2  J_\lambda \Pi_{\lambda\mu} \Gamma_{\mu\nu} \Pi_{\rho\nu}
   J_\rho
 \ri \} \ri \}
\label{z4}
\ee
where 
$\Gamma_{\mu\nu} = [ \delta_{\mu\nu} - \lambda^2 \Pi_{\mu\nu} ]^{-1}
  = \delta_{\mu\nu} + O(\lambda^2) $. Since we are
working up to the second order in $\lambda$, therefore, we can just
retain the $\delta_{\mu\nu}$ piece in (\ref{z4}).

The two-point current correlation function is easily obtained from
(\ref{z4})
\be
-\fr{1}{\lambda^2}
\fr{\delta^2}{\delta J_\mu (q) \delta J_\nu (-q)}|_{J=0} =
<j_\mu (q) j_\nu (-q) > = 
 \Pi_{\mu\nu}(q)
+ \lambda^2  \Pi_{\mu\alpha} (q) \Pi_{\alpha\nu} (q)
\label{jj}
\ee
Now, it is not difficult to extract the bosonized lagrangean
of the MTM in this limit. This is given by
\be
\cl_{MTM} =  \fr{1}{2} B_\mu 
 \lef (
 \Pi_{\mu\nu}
+ \lambda^2  \Pi_{\mu\alpha} \Pi_{\alpha\nu} \ri) B_\nu
\label{bl}
\ee
Observe that 
this is a gauge theory, because of the transverse nature
of the kernel. It is nonlocal for any value of the fermion
mass except for $m \rightarrow \infinity$ as can be explicitly checked
from expressions (\ref{1a}) and (\ref{b}). This is a general feature
of bosonization in higher dimensions \cite{em1,bfo,bm,bm2} occurring 
even in the
free case.
It has been shown that 
generalized free (quadratic) gauge theories of this type,
in spite of being nonlocal, 
yield sensible results \cite{bfo,bm,ma}.
In particular, they respect causality.

It is now easy to deduce the current bosonization formula
which will reproduce the correlation function (\ref{jj}). This is given
by
\be
j_\mu =  \lef (  \Pi_{\mu\nu} 
+ \lambda^2  \Pi_{\mu\alpha} \Pi_{\alpha\nu} \ri) B_\nu
\label{j}
\ee
To see that this is correct, we must consider the 
two-point correlation funtion of the $B_\mu$-field which is given by the
inverse of the kernel appearing in the quadratic lagrangean (\ref{bl}).
In order to perform this inversion we add
a gauge fixing term $\xi B(q^2) q_\mu q_\nu$.
As it turns out it will also be convenient to add the same longitudinal term
to the second $\Pi_{\mu\nu}$
in the $\lambda$-dependent term of (\ref{bl}) which is allowed by
the transversality of $\Pi_{\mu\nu}$.
The result for the field two-point function in this gauge is
$$
<B_\mu(q) B_\nu (-q) > = 
\lef [ \Pi_{\mu\alpha} (q)\lef ( \delta_{\alpha\nu} +
\lambda^2 \lef ( \Pi_{\alpha\nu} (q) +
\xi B(q^2)  q_\alpha q_\nu \ri ) \ri) +
\xi B(q^2) q_\mu q_\nu  \ri ]^{-1} 
$$
\be
= D_{\mu\nu} (q)
- \lambda^2 \fr{P_{\mu\nu}(q)}{q^2} + O(\lambda^4)
\label{bb}
\ee
where
$$
D_{\mu\nu}(q) =
\lef [\Pi_{\mu\nu} (q) + \xi B(q^2) q_\mu q_\nu \ri ]^{-1} 
$$
\be
= \fr{1}{q^2 [A^2 (q^2) + q^2 B^2 (q^2) ] }
\lef [ B(q^2) P_{\mu\nu} - A(q^2) C_{\mu\nu}
\ri ] + \fr{1}{\xi} \fr{q_\mu q_\nu }{q^4 B (q^2)}
\label{d}
\ee
Now, using (\ref{j}), we have
\be
<j_\mu (q) j_\nu (-q) > = 
 \lef (  \Pi_{\mu\alpha} 
+ \lambda^2  \Pi_{\mu\beta} \Pi_{\beta\alpha} \ri) (q)
 \lef (  \Pi_{\nu\rho} 
+ \lambda^2  \Pi_{\nu\sigma} \Pi_{\sigma\rho} \ri) (-q)
< B_\alpha (q) B_\rho (-q) >
\label{jj1}
\ee
It is straightforward to see that the current correlation function
(\ref{jj}) is reproduced by inserting (\ref{bb}) in (\ref{jj1}).
Also, it is not difficult to prove that all higher correlation functions
are reproduced by the bosonic expression (\ref{j}).
Note that,
in particular, the odd functions vanish because the odd $B_\mu$-correlators
are zero in the quadratic gauge field theory (\ref{bl}).
The current bosonization formula (\ref{j}) is thereby confirmed,
also showing that in this limit the MTM is mapped into a generalized free
gauge theory (\ref{bl}).

Similarly to the two-dimensional case, we can now define 
a dual current as
\be
\bar j_\mu (q) = \fr{i}{q} C_{\mu\alpha} (q) j_\alpha (q)
\label{jb}
\ee
The correlation functions of this dual current are identical to those
of $j_\mu$ as can be be verified from the bosonized expression (\ref{j}).
This generalizes a similar duality relation found in the large mass
limit \cite{rb}.

Let us next consider the exact bosonization of the free massive fermionic
theory which is obtained in the limit when the Thirring coupling
$\lambda$ vanishes. From (\ref{bl}) and (\ref{j})
we immediately obtain the bosonization
formulae for the lagrangean and current, namely
\be
\bar\psi (- i \not\! q + m ) \psi |_{\rm free}
=
  \fr{1}{2} B_\mu  \Pi_{\mu\nu}  B_\nu
\label{fb}
\ee
\be
j_\mu (q) |_{\rm free} =   \Pi_{\mu\nu} B_\nu
\label{fb1}
\ee
For a vanishing mass, these expressions reduce exactly to
the ones found by following a direct operator bosonization of the
free massless fermion field \cite{em1}. This clearly shows that the
operator realization obtained in \cite{em1} is exact and no nonquadratic
corrections are necessary.

A very important point must be stressed now.
These exact bosonization formulae
are strictly valid only in the free case. In an interacting theory,
both expressions are modified in general. The current,
for instance, is given by
(\ref{j}), while the kinetic fermion lagrangian for the MTM 
can be obtained from (\ref{bl}) and (\ref{j}) giving the result
\be
\bar\psi (- i \not\! q + m ) \psi |_{\rm MTM} =
\bar\psi (- i \not\! q + m ) \psi |_{\rm free} +\frac
{\lambda^2}{2}   B_\mu(q)\Bigl( \Pi_{\mu\alpha}  \Pi_{\alpha\nu}(q) 
+ \Pi_{\mu\alpha}  \Pi_{\alpha\nu}(-q)\Bigr)B_\nu(-q)
\label{fb111}
\ee
This relation clearly shows the modification of the bosonization formula
for the kinetic fermion lagrangian which is produced 
when we add the Thirring interaction to the free massive lagrangian. Self-
consistency is preserved as can be inferred by setting $\lambda = 0$
in the above formula. The fact that the bosonization formulae are
changed in the presence of an interaction is a general feature of
dimensions higher than two \cite{bm}.

Our results demonstrate that, because of the finiteness
of the one-loop vacuum polarization tensor,
even though the MTM is pertubatively
nonrenormalizable by the usual power counting criterion,
at least in the weak coupling regime, it yields sensible results.

\subsection{ Master Lagrangean Approach }

This method starts by considering the theory of a fermion interacting
with a vector gauge field which, by its turn, has a topological
interaction with a dynamical Maxwell field \cite{rb}.
The corresponding lagrangean is called Master Lagrangean
since integrating out the gauge fields leads to the MTM, while the
bosonized form is obtained by integrating over the fermion field.
It is given by \cite{rb}
\be
\cl_{M} = \bar\psi (- \not\!\del + m +i \lambda \not\!\! f ) \psi
- \fr{1}{4} F_{\mu\nu} F_{\mu\nu}
- i \epsilon_{\mu\nu\lambda}
f_\mu \del_\nu A_\lambda
\label{ml}
\ee
where $F_{\mu\nu} = \del_\mu A_\nu - \del_\nu A_\mu$. The corresponding
current correlators generating functional in euclidean space is given by
\be
Z[J] = \int  D\psi D\bar\psi D f_\mu D A_\mu
\delta(\del_\mu f_\mu) \delta( \del_\mu A_\mu) 
\exp \lef\{-\int d^3z \lef [ 
\cl_{M}
+ i \epsilon_{\mu\nu\lambda}
J_\mu \del_\nu A_\lambda \ri ]  \ri \}
\label{z5}
\ee
where a covariant gauge fixing has been chosen for both the fields.
It is easily seen by integrating out the gauge fields that the MTM
current generating functional (\ref{z1}) is reproduced \cite{rb}.
From this observation we can immediately conclude that the exact
bosonization formula in momentum space for the Thirring current
in terms of the $A_\mu$-field is given by 
\be
j_\mu = - \fr{i}{\lambda} C_{\mu\nu}  A_\nu
\label{j1}
\ee
It is important to note at this point that the above result is exact
and in particular does not depend on 
any weak coupling approximation in the Thirring coupling.
Moreover, this is the only bosonization formula in dimensions
higher than two which does not depend on the type of interacting
theory in which the bosonized operator is embedded.
We can see, for instance, that a
completely different result is obtained 
through the current correlators method
where both the lagrangean and current are modified as can be seen
from (\ref{bl}) and (\ref{j}).
Let us stress that
the vector gauge fields $B_\mu$ and $A_\mu$ which appear in the
two bosonization formulas (\ref{j}) and (\ref{j1}) of course are not the
same. Later on, we will explicitly show the equivalence of the two
descriptions and find out the relationship between these fields.

In order to obtain the bosonized form of the MTM lagrangean, let us
integrate over the fermion fields. This will be done as before,
by means of a weak coupling expansion. We find the generating functional
$$
Z[J] = \int   D f_\mu D A_\mu
\delta(\del_\mu f_\mu) \delta( \del_\mu A_\mu) 
\exp \Bigl \{-\int d^3z \Bigl [ 
-\fr{1}{4} F_{\mu\nu} F_{\mu\nu}
\Bigr . \Bigr .
$$
\be
\Bigl . \Bigl .
+ \fr{\lambda^2 }{2} f_\mu \Pi_{\mu\nu} f_\nu 
- i \epsilon_{\mu\nu\lambda}(f_\mu -
J_\mu) \del_\nu A_\lambda\Bigr ] \Bigr \}
\label{z6}
\ee
where $\Pi_{\mu\nu}$ is the one-loop vacuum polarization tensor given
in momentum space by (\ref{pi}). Now, performing the quadratic functional
integration over $f_\mu$ we get
\be
Z[J] = \int   D A_\mu
\delta( \del_\mu A_\mu) 
\exp \lef\{-\int d^3z \lef [ 
\fr{1}{2} A_\mu \Sigma_{\mu\nu} A_\nu
+ i \epsilon_{\mu\nu\lambda}
J_\mu \del_\nu A_\lambda \ri ] \ri \}
\label{z7}
\ee
where 
\be
\Sigma_{\mu\nu} =  \lef (
 \fr{B(q^2)}{\lambda^2 (A^2 (q^2) + q^2 B^2 (q^2) )}-1 \ri ) P_{\mu\nu} -
\fr{A(q^2)}{\lambda^2 (A^2 (q^2) + q^2 B^2 (q^2) )} C_{\mu\nu} 
\label{si}
\ee
The complete bosonization of the MTM lagrangean in the weak
coupling limit is therefore given by
\be
\cl_{MTM} = 
\fr{1}{2} A_\mu \Sigma_{\mu\nu} A_\nu
\label{bl1}  
\ee
which, by the transversality of $\Sigma_{\mu\nu}$, is also a
generalized free gauge theory.
Note that just as the bosonization of the current in
terms of the $A_\mu$-field (\ref{j1}) was shown to differ from that
obtained by the previous method, the same happens for the corresponding
bosonized lagrangeans. However, as we now show, the current correlation
functions derived in the present formalism are identical to
(\ref{jj}), which was also reproduced by the previous bosonization
formula (\ref{j}) in terms of the $B_\mu$-field. Indeed, performing the
$A_\mu$-field integration in (\ref{z7}) and keeping terms up to the
order $O(\lambda^2)$, we obtain 
\be
Z[J] =    
\exp \lef\{-\int d^3 q 
 \fr{\lambda^2}{2} J_\mu (q) \lef [
 \Pi_{\mu\nu}(q)
+ \lambda^2  \Pi_{\mu\alpha} (q) \Pi_{\alpha\nu} (q)
\ri ] J_\nu (-q) \ri \}
\label{z8}
\ee
Now, it is straightforward to see that by taking functional derivatives
of (\ref{z8}) with respect to the sources the two-point correlation
function (\ref{jj}) is reproduced. Similarly, all the higher
current correlators can be obtained as well. As usual, the odd ones
will vanish.

We have shown how the two distinct methods of bosonization considered
here can give equivalent descriptions of the same fermionic theory. It
is also possible to provide a direct mapping between the basic bosonic
fields in the different approaches, namely, the $A_\mu$ and
$B_\mu$ fields. Any divergenceless field in three dimensions like $A_\mu$ in
the covariant gauge we are working can be expressed as
\be
A_\mu = \lef [ a(q^2 ) P_{\mu\alpha} + b(q^2) C_{\mu\alpha} \ri ]
B_\alpha
\label{a}
\ee
where $a(q^2 )$ and $b(q^2 )$ are scalar functions and $B_\mu$ is some
other vector field. Identifying $A_\mu$ and $B_\mu$, respectively, with
our two bosonic gauge fields and inserting (\ref{a}) in (\ref{j1}), we
can determine the unknown functions $a$ and $b$ by comparing the
resulting expression with (\ref{j}). We find
$$
a(q^2) = i\lambda \lef [ \fr{A(q^2)}{q^2} + 2 \lambda^2 A(q^2) B(q^2)
\ri ]
$$
\be
b(q^2) = i\lambda \lef [- B(q^2) +  \lambda^2 ( A^2(q^2) - q^2 B^2 (q^2) )
\ri ]
\label{ab}
\ee
It can now be verified that inserting (\ref{a}) with (\ref{ab}) in the
bosonic lagrangean of the MTM, eq. (\ref{bl1}), we precisely
reobtain the alternative form given in (\ref{bl}). This establishes
the complete equivalence of the two bosonization methods.

We now discuss the inverse of the mapping expressed by (\ref{a}). The
operator multiplying $B_\mu$ in (\ref{a}) is not invertible
because of transversality. Hence, taking advantage of the gauge condition
satisfied by $B_\mu$, we can add an extra longitudinal piece as we did
in (\ref{bb}), to obtain
\be
A_\mu = \lef [ a(q^2 ) P_{\mu\alpha} + b(q^2) C_{\mu\alpha}
+ \xi B(q^2) q_\mu q_\alpha \ri ]
B_\alpha
\label{2a}
\ee
We can now invert the relevant operator and get
\be
B_\mu =
\fr{1}{q^2 ( b^2 (q^2) + q^2 a^2 (q^2) )}
\lef [ a(q^2 ) P_{\mu\alpha} - b(q^2) C_{\mu\alpha} \ri ]
A_\alpha
\label{3a}
\ee
where explicit use has been made of the covariant gauge condition satisfied
by $A_\mu$. It can be shown that using (\ref{3a}),
in the bosonized expressions for the lagrangian and
current given respectively by (\ref{bl}) and (\ref{j}), we reproduce
the corresponding ones in terms of $A_\mu$, namely (\ref{bl1}) and
(\ref{j1}). This establishes the algebraic consistency of the whole
program.

We conclude the section by giving the exact explicit expressions for the
lagrangean and the current for a free massive fermion, 
which follow from (\ref{bl1}) and (\ref{j1}),
respectively, by making the
scaling $A_\mu \rightarrow \lambda A_\mu$ and then setting $\lambda =0$,
namely
\be
\bar\psi (- i \not\! q + m ) \psi |_{\rm free}
=
  \fr{1}{2} A_\mu  \Sigma^f_{\mu\nu}  A_\nu
\label{x1}
\ee
\be
j_\mu (q) |_{\rm free} = -i C_{\mu\nu} A_\nu
\label{x2}
\ee
where
\be
\Sigma^f_{\mu\nu} =  
 \fr{B(q^2)}{A^2 (q^2) + q^2 B^2 (q^2) }  P_{\mu\nu} -
\fr{A(q^2)}{A^2 (q^2) + q^2 B^2 (q^2) } C_{\mu\nu} 
\label{sif}
\ee
Note that the above equations are the analogs of (\ref{fb}) and (\ref{fb1})
obtained in the current correlator approach. 
Observe also that the current has the
general topological structure as (\ref{j1}).

As in the interacting theory, the fields $A_\mu$ and $B_\mu$ are
related by an expression identical to (\ref{a}) except for the fact that
the scalar functions (\ref{ab}) are replaced by
$$
a^f (q^2) =  i \fr{A(q^2)}{q^2} 
$$
\be
b^f (q^2) = -i  B(q^2) 
\label{abf}
\ee

Just as in the previous approach, the results (\ref{x1}) and (\ref{x2})
are strictly exact only in the free theory.

\section{Massive Thirring Model in Four Dimensions}

Let us now consider the bosonization of the MTM in four dimensions
by following
the two methods applied in the previous section. An important difference
from the three-dimensional case is that now the one-loop vacuum
polarization tensor is no longer finite and renormalization is necessary.
The implications of this will be analyzed.

\subsection{ Current Correlator Approach}
\setcounter{equation}{0}

Again we start from the current correlators generating functional
in four dimensional euclidean space
\be
Z[J] = \int  D\psi D\bar\psi
\exp \lef\{-\int d^4z \lef [ 
\bar\psi (- \not\!\del + m ) \psi - \fr{\lambda^2}{2} j_\mu j_\mu +
i\lambda j_\mu  J_\mu \ri ] \ri \}
\label{Z1}
\ee
As usual this can be written as
\be
Z[J] = \int  D\psi D\bar\psi D A_\mu
\exp \lef\{-\int d^4z \lef [ 
\bar\psi (- \not\!\del + m +i e\  \not\!\!\! A ) \psi
- \fr{\mu^2}{2} A_\mu A_\mu +
i\lambda j_\mu  J_\mu \ri ] \ri \}
\label{Z2}
\ee
where $\lambda = \fr{e}{\mu}$.
We now integrate over the fermion field in the small coupling approximation
where only the two-legs one-loop graph does contribute. We obtain
\be
Z[J] = \int   D A_\mu
\exp \lef\{-\int d^4z \lef \{ 
 \fr{ e_R^2}{2} ( A_\mu + \fr{J_\mu}{\mu} ) \Pi_{\mu\nu} ( A_\nu
+ \fr{J_\nu )}{\mu}
- \fr{\mu^2}{2} A_\mu A_\mu 
 \ri \} \ri \}
\label{Z3}
\ee
where the vacuum polarization tensor is given by \cite{iz}
$$
\Pi_{\mu\nu}(q) = (q^2 \delta_{\mu\nu}- q_\mu q_\nu) \Pi(q^2)
= P_{\mu\nu} \Pi(q^2)
$$
with
\be
\Pi(q^2)= -\fr{1}{12\pi^2 }\lef \{ \fr{1}{3} + 2
\lef (1-\fr{2m^2}{q^2}\ri ) \lef [ \fr{1}{2} x \ln \fr{x+1}{x-1} - 1 \ri]
\ri\}
\label{Pi}
\ee
in which $x=\lef (1+\fr{4m^2}{q^2}\ri )^{1/2}$. In the above expression, the
renormalized coupling constant $e_R$ is given, in lowest order, by
\be
e_R^2 = \lef [ 1 - \fr{e_R^2}{12\pi^2} \ln \Lambda^2
\ri ] e^2
\label{lam}
\ee
where $\Lambda$ is an ultraviolet cutoff.

At this point, let us remark that the coupling constant $e$ is
renormalized exctly as in QED because the evaluation of the fermionic
determinant in (\ref{Z2}) was performed in the renormalizable sector
of the effective theory defined by the action in (\ref{Z2}). In this
sense it is meaningful to use a small coupling expansion in $e_R$.
On the other hand, by rescaling $A_\mu \rightarrow \mu A_\mu$, it
is possible to redefine the coupling as $\fr{e_R}{\mu}$. Consistency
then requires that an expansion in small $e_R$ must be supplemented
by the requirement that $e_R << \mu$. Note that the original MTM
with the coupling $\lambda = \fr{e}{\mu}$ is nonrenormalizable. This
discussion clarifies the precise meaning of a small coupling expansion
in MTM: small coupling  expansion in the effective theory implies the
corresponding expansion in the MTM with a 
redefined coupling $\tilde \lambda = \fr{e_R}{\mu}$. For notational
convenience we henceforth set $\mu = 1$, so that $\tilde \lambda = e_R$.
Observe that in the three-dimensional case the one-loop vacuum
polarization tensor was finite and therefore no renormalization was
necessary. Hence there was no need to introduce a redefined Thirring
coupling.

We can now perform the quadratic integration over $A_\mu$ in (\ref{Z3})
by using the fact that up to the order $O(\lt^2)$ the inverse of the
kernel
\be
\lef [ \delta_{\mu\nu} - \lt^2  \Pi_{\mu\nu} \ri ]^{- 1} =
\delta_{\mu\nu} + O(\lt^2)
\label{ik}
\ee
The result is
\be
Z[J] = 
\exp \lef\{-\int d^4z  \fr{ \lt^2}{2} \lef \{ 
   J_\mu  \Pi_{\mu\nu}  J_\nu 
+ \lt^2  J_\lambda \Pi_{\lambda\mu}  \Pi_{\rho\mu}
   J_\rho
 \ri \} \ri \}
\label{Z4}
\ee

Now, taking functional derivatives of the above expression with respect to
the sources, we immediately find the two-point current correlation function
\be
 -\fr{1}{\lt^2} \fr{\delta^2}{\delta J_\mu (q) \delta J_\nu (-q)}|_{J=0} =
  <j_\mu (q) j_\nu (-q) > = 
 \Pi_{\mu\nu}(q)
+ \lt^2  \Pi_{\mu\alpha} (q) \Pi_{\alpha\nu} (q)
\label{JJ}
\ee
The bosonized form of the MTM lagrangian in the small $\lt$ regime
is now easily inferred, namely
$$
\cl_{MTM} =
\fr{1}{2} B_\mu
 \lef (
 \Pi_{\mu\nu}
+ \lt^2  \Pi_{\mu\alpha} \Pi_{\alpha\nu} \ri) B_\nu
$$
\be
=
\fr{1}{2} B_\mu
\Pi (q^2)
 \lef (
1 + \lt^2  q^2  \Pi (q^2) \ri)   P_{\mu\nu} B_\nu
\label{BL}
\ee
The simplification in the second line happens because the vacuum
polarization tensor has only the $P_{\mu\nu}$ part in four
dimensions.
Note that this bosonized form of the lagrangian is structurally identical
to the one obtained for the corresponding bosonization formula 
(\ref{bl}) in three dimensions.
This also shows the practical viability of the
current correlator approach for bosonization.

In the same way as in three dimensions,
we now proceed to the bosonization of the
current. This can be immediately obtained by inspecting the previous
formula, given by (\ref{j}), namely
$$
j_\mu = 
 \lef (  \Pi_{\mu\nu} 
+ \lt^2  \Pi_{\mu\alpha} \Pi_{\alpha\nu} \ri) B_\nu
$$
\be
=
 \Pi (q^2)
 \lef (
1 + \lt^2  q^2  \Pi (q^2) \ri)   P_{\mu\nu} B_\nu
\label{J}
\ee
Following exactly the same steps as in the previous section, we can show
that the above bosonized expression for $j_\mu$ precisely reproduces
the two-point correlator (\ref{JJ}), as well as the higher ones.
The odd funtions as usual vanish. This confirms the validity of our
bosonization procedure also in this case.

The free case can now be easily obtained as in the three-dimensional 
theory, by taking the limit $\lt \rightarrow 0$. We obtain
the exact identifications
\be
\bar\psi (- i \not\! q + m ) \psi |_{\rm free}
=
  \fr{1}{2} B_\mu  \Pi_{\mu\nu}  B_\nu =
  \fr{1}{2} B_\mu \Pi (q^2) P_{\mu\nu}  B_\nu
\label{fbf}
\ee
and
\be
j_\mu (q) |_{\rm free} =   \Pi_{\mu\nu} B_\nu =
  \Pi (q^2) P_{\mu\nu}  B_\nu
\label{fbf1}
\ee
Let us remind again that in the presence of interaction both the
above expressions are modified and, in particular, a relation
exactly analogous to
(\ref{fb111}) is obtained.

\subsection{ Master Lagrangean Approach }

Let us now apply the Master Lagrangean method to the MTM in four
dimensions. In this case, we have to consider the
theory of a fermion coupled to a vector gauge field which topologically
interacts with a dynamical Kalb-Ramond
second rank antisymmetric tensor gauge field
\cite{rb1,bm}.
\be
\cl_{M} = \bar\psi (- \not\!\del + m +i \lambda \not\!\! f ) \psi
- \fr{1}{3} F_{\mu\nu\lambda} F_{\mu\nu\lambda}
- i \epsilon_{\mu\nu\alpha\beta}
f_\mu \del_\nu A_{\alpha\beta}
\label{ML}
\ee
where $F_{\mu\nu\lambda} = \del_\mu A_{\nu\lambda}
- \del_\nu A_{\mu\lambda} - \del_\lambda A_{\nu\mu}$.
The euclidean generating functional of current correlators associated
with this is given by
\be
Z[J] = \int  D\psi D\bar\psi D f_\mu D A_{\mu\nu} D \alpha
\delta(\del_\mu f_\mu) \delta( \del_\mu A_{\mu\nu} + \del_\nu \alpha) 
\exp \lef\{-\int d^4z \lef [ 
\cl_{M}
+ i \epsilon_{\mu\nu\alpha\beta}
 \del_\nu A_{\alpha\beta}  J_\mu \ri ]  \ri \}
\label{zz5}
\ee
Notice that the delta-functional in the Kalb-Ramond field
includes an additional term which is integrated on. This
accounts for the reducibility in the
usual covariant gauge fixing which is related to the fact that the
gauge invariance in the Kalb-Ramond field is reducible.
It can be easily verified that upon integration over the gauge fields we
reproduce the MTM current generating functional (\ref{Z1}). From this
again we can immediately infer the exact bosonization formula for the
current, namely
\be
j_\mu =  \fr{i}{\lambda} \epsilon_{\mu\nu\alpha\beta}q_\nu
A_{\alpha\beta}
\label{jj11}
\ee
Note that in analogy with the three-dimensional result (\ref{j1}),
the bosonized current is exactly given by the topological current in
terms of the Kalb-Ramond field. The previous comments  made below
eq. (\ref{j1}) regarding the comparison between the two bosonization
methods of the fermion current also apply here.

The bosonized form of the MTM lagrangean is now obtained  from (\ref{zz5})
by integrating out the fermion fields. The result in the weak coupling
approximation is given by,
$$
Z[J] = \int   D f_\mu D A_{\mu\nu} D\alpha
\delta(\del_\mu f_\mu) \delta( \del_\mu A_{\mu\nu} + \del_\nu \alpha) 
\exp \lef\{-\int d^4z \lef [
- \fr{1}{3} F_{\mu\nu\lambda} F_{\mu\nu\lambda}
 \ri . \ri .
$$
\be
\lef . \lef .
- i \epsilon_{\mu\nu\alpha\beta}
f_\mu \del_\nu A_\alpha\beta
+ \fr{\lt^2}{2} f_\mu \Pi_{\mu\nu} f_\nu
+ i \epsilon_{\mu\nu\alpha\beta}
\del_\nu A_{\alpha\beta}  J_\mu
\ri ] \ri \}
\label{zz6}
\ee
where $\Pi_{\mu\nu}$ is the one-loop vacuum polarization tensor
already given
in momentum space by (\ref{Pi}).
Observe that, as explained before, the original MTM coupling $\lambda$
is replaced by the redefined coupling $\lt$.
Next, performing the quadratic functional
integration over $f_\mu$ we obtain
$$
Z[J] = \int   D A_{\mu\nu} D\alpha
 \delta( \del_\mu A_{\mu\nu} + \del_\nu \alpha) 
\exp \Bigl\{-\int d^4z \Bigl [
 \fr{1}{3} F_{\mu\nu\lambda}
\Bigl [ \fr{1}{q^2 \lt^2  \Pi (q^2)}-1 \Bigr ]
F_{\mu\nu\lambda} \Bigr .\Bigr .
$$
\be
\Bigl . \Bigl .
+ i \epsilon_{\mu\nu\alpha\beta}
\del_\nu A_{\alpha\beta}  J_\mu
\Bigr ] \Bigr \}
\label{1zz7}
\ee
from which we can read the explict bosonization expression for the
MTM lagrangean in the weak coupling limit and arbitrary mass, in terms
of the second rank antisymmetric Kalb-Ramond field
\be
\cl_{MTM} =
 \fr{1}{3} F_{\mu\nu\lambda}
\lef [ \fr{1}{q^2 \lt^2  \Pi (q^2)}-1 \ri ]
F_{\mu\nu\lambda}
\label{BL1}
\ee
 Notice that in the weak coupling regime the first piece dominates leading
to a positive definite lagrangean, as expected in a euclidean metric. The
same also holds for the lagrangean (\ref{bl1}) in the three dimensional
case. Positive definiteness of the bosonised lagrangians obtained
by the current correlator method is self evident, as is easily seen by looking
at the respective expressions.

In the same way
as has happened in the case of the current, notice that the bosonized
lagrangian (\ref{BL1}) 
also differs from the one obtained by the previous method,
eq. (\ref{BL}). However, as we now explicitly show, the same fermionic
current correlation functions (\ref{JJ}) are reproduced. This can be easily
seen by integrating over $\alpha$ and the Kalb-Ramond field in (\ref{1zz7}),
\be
Z[J] = 
\exp \lef\{-\int d^4z 
 \fr{\lt^2}{2} J_{\mu} \lef [
 \Pi_{\mu\nu}(q)
+ \lt^2  \Pi_{\mu\alpha} (q) \Pi_{\alpha\nu} (q)
\ri ]
J_{\nu}
 \ri \}
\label{x3}
\ee
and taking functional derivatives.

We shall now explicitly derive the relationship between the Kalb-Ramond
field and the vector gauge field used in the current
correlator approach, thereby
generalizing the relation (\ref{a}). Note that in the covariant gauge, any
Kalb-Ramond field can always be written in terms of a vector field as
\be
A_{\mu\nu} = g(q^2) \epsilon_{\mu\nu\alpha\beta} q_\alpha B_\beta
\label{amn}
\ee
where $g(q^2)$ is some scalar function. By identifying the above gauge
fields with those occurring in the two bosonization approaches cosidered
above, it is possible to determine the scalar function by the direct
comparison of the two bosonization formulas for the current (\ref{J})
and (\ref{jj11}). We find
\be
 g(q^2) =  i \fr{\lt}{2} 
 \Pi (q^2)
 \lef (
1 + \lt^2  q^2  \Pi (q^2) \ri)
\label{f}
\ee
With this mapping, it is simple to check that the bosonized lagrangeans
obtained in the two different approaches, namely (\ref{BL}) 
and (\ref{BL1}) also become identical. 
Following the steps of the
previous section we can invert relation (\ref{amn}) and thereby
demonstrate the complete
equivalence between the two approaches in both directions.

It is now simple to read off the exact results for the free massive theory
by scaling $A_{\mu\nu} \rightarrow \lt A_{\mu\nu}$
and finally making $\lt =0$:
\be
\bar\psi (- i \not\! q + m ) \psi |_{\rm free}
=
\fr{1}{3} F_{\mu\nu\lambda}
\lef [  \fr{1}{q^2   \Pi (q^2)} \ri ]
F_{\mu\nu\lambda}
\label{fbf1}
\ee
\be
j_\mu (q) |_{\rm free} =
i\epsilon_{\mu\nu\alpha\beta}q_\nu
A_{\alpha\beta}
\label{fbf11}
\ee
The relation mapping the two fields remains identical to (\ref{amn}),
except for the fact that the scalar function $g(q^2)$ is modified
as
\be
g^f(q^2) =  \fr{i}{2}
 \Pi (q^2)
\label{f1}
\ee

\section{ The Two-Dimensional Theory Revisited}

\setcounter{equation}{0}

Let us consider here the application of the two methods discussed
in the previous sections to the bosonization of the two-dimensional
massless Thirring Model which is known to be exactly solvable.
The euclidean current correlator generating functional is given by
\be
Z[J] = \int  D\psi D\bar\psi
\exp \lef\{-\int d^2 z \lef [ 
\bar\psi (- \not\!\del ) \psi - \fr{e^2}{2} j_\mu j_\mu +
i e \  j_\mu  J_\mu \ri ] \ri \}
\label{Z11}
\ee
As done previously we can eliminate the four-fermion interaction
by introducing a vector field, namely
\be
Z[J] = \int  D\psi D\bar\psi D A_\mu
\exp \lef\{- \int d^2z \lef [ 
\bar\psi (- \not\!\del +i e\  \not\!\!\! A ) \psi
- \fr{1}{2} A_\mu A_\mu +
i e \  j_\mu  J_\mu \ri ] \ri \}
\label{Z22}
\ee
The fermion integration can be done exactly giving the result
\be
Z[J] = \int   D A_\mu \delta(\del_\mu A_\mu)
\exp \lef\{-\int d^2z \lef [ 
 \fr{ e^2}{2\pi} ( A_\mu + J_\mu ) \Gamma_{\mu\nu} ( A_\nu
+ J_\nu )
- \fr{1}{2} A_\mu A_\mu 
 \ri ] \ri \}
\label{Z33}
\ee
where
\be
\Gamma_{\mu\nu} = \delta_{\mu\nu} - \fr{\del_\mu \del_\nu}{\Box}
\label{gama}
\ee
Performing the gaussian integration in (\ref{Z33}), we get
\be
Z[J] = 
\exp \lef\{-\int d^2 z \lef ( \fr{ e^2}{2 ( \pi - e^2 )}  \ri )
   J_\mu  \Gamma_{\mu\nu}  J_\nu 
 \ri \} 
\label{Z44}
\ee
It is now trivial to compute the current two-point correlation function
by taking functional derivatives with respect to $J_\mu$. The result
is
\be
- \fr{1}{e^2} \fr{\delta^2}{\delta J_\mu (x) \delta J_\nu (y)}|_{J=0} =
  <j_\mu (x) j_\nu (y) > =
\lef (  \fr{1}{ ( \pi - e^2 )}\ri )  \lef [ \delta_{\mu\nu} \delta^2 (x-y)
  - \del^x_\mu \del^x_\nu \lef[ \fr{1}{\Box}\ri ] (x-y)\ri ]
\label{JJ11}
\ee
As done earlier, the bosonized form of the lagrangian is immediately
inferred from this result,
\be
\cl_{Thirring} =
\lef ( \fr{1}{2 ( \pi - e^2 )}\ri ) B_\mu
\Gamma_{\mu\nu} B_\nu
\label{bl22}
\ee
Similarly, the bosonization of the current yields
\be
j_\mu = \lef ( \fr{1}{\pi - e^2} \ri ) \Gamma_{\mu\nu} B_\nu
\label{j22}
\ee
Now it is simple to verify the validity of these bosonization rules
by evaluating the current correlation function in the bosonic language.
In order to do it, we
obtain the propagator of the $B_\mu$-field from (\ref{bl22})
in some covariant gauge
and using it in (\ref{j22}), easily reproduce (\ref{JJ11}).

The results for the free theory follow by simply putting $e=0$ in
(\ref{bl22}) and (\ref{j22}). This clearly shows why the Thirring
model is a free theory since the free and interacting cases just
differ by a normalization of the $B_\mu$-field. Observe the distinction
from the higher dimensional situation, where the connection between
the free and interacting cases is highly nontrivial.

One may wonder how to relate this bosonization result to the usual
one in which the bosonic field is scalar. This will be done by
adopting the Master Lagrangean approach \cite{rb1} where the
corresponding lagrangean
is now defined by
\be
\cl_{M} = \bar\psi (- \not\!\del  +i e \not\!\! A ) \psi
- \fr{1}{2} \del_{\mu} \phi  \del_{\mu} \phi
- i \epsilon_{\mu\nu}
A_\mu \del_\nu \phi
\label{ML11}
\ee
As before, the current correlators generating functional is given by
\be
Z[J] = \int  D\psi D\bar\psi D \phi D A_{\mu} 
\delta(\del_\mu A_\mu) 
\exp \lef\{-\int d^2z \lef [ 
\cl_{M}
+ i \epsilon_{\mu\nu}
 \del_\nu \phi J_\mu \ri ]  \ri \}
\label{zz55}
\ee
Doing the integration over the bosonic fields, one immediately
reproduces the current correlator generating functional of the massless
Thirring model, given by (\ref{Z11}). From this we infer the exact
bosonization formula for the Thirring current,
\be
j_\mu = \fr{1}{e} \epsilon_{\mu\nu} \del_\nu \phi
\label{1}
\ee
Note that this is the well known form given in the literature.

Alternatively, doing the fermionic integration in (\ref{zz55}), we obtain
\be
Z[J] = \int D\phi   D A_\mu \delta(\del_\mu A_\mu)
\exp \lef\{- \int d^2z \lef [ 
 \fr{ e^2}{2\pi}  A_\mu  \Gamma_{\mu\nu}  A_\nu
- i \epsilon_{\mu\nu}
A_\mu  \del_\nu  \phi
- \fr{1}{2} \del_{\mu} \phi  \del_{\mu} \phi
+ i \epsilon_{\mu\nu}
 \del_\nu \phi J_\mu 
\ri ]  \ri \}
\label{Z333}
\ee
Doing the $A_\mu$ integration leads to
\be
Z[J] = \int D\phi  
\exp \lef\{- \int d^2z \lef [ 
\lef ( \fr{\pi - e^2}{2 e^2} \ri )
 \del_{\mu} \phi  \del_{\mu} \phi
+ i \epsilon_{\mu\nu}
 \del_\nu \phi J_\mu 
\ri ]  \ri \}
\label{Z3333}
\ee
From this we see that the exact bosonization of the Massless Thirring
Model in terms of the scalar field is given by
\be
\cl_{Thirring} =
\lef ( \fr{\pi - e^2}{2 e^2} \ri )
\del_{\mu} \phi  \del_{\mu} \phi
\label{3}
\ee
A simple scaling reproduces the well knowm lagrangean for a free massless
scalar field.

We finally show the equivalence between the conventional expressions
and those given earlier in terms of the vector gauge field.
Indeed, performing the integration over $\phi$ in (\ref{Z3333}) yields
\be
Z[J] = 
\exp \lef\{- \int d^2z 
\lef ( \fr{e^2}{2(\pi - e^2)} \ri )
 J_\mu  \Gamma_{\mu\nu}  J_\nu
  \ri \}
\label{2}
\ee
from which one easily reproduces the current correlation functions
found earlier (\ref{JJ11}).

It is now straightforward to verify that the mapping between the
bosonic vector and scalar fields in the two bosonization schemes is
given by
\be
B_\mu =  \lef (\fr{\pi - e^2 }{e} \ri ) \epsilon_{\mu\nu} \del_\nu \phi
\label{4}
\ee
This is also expected on general grounds since in two dimensions any
vector field in a transverse gauge can always be expressed in terms
of the curl of a scalar field.

\section{Conclusions}

Two different approaches to bosonization were fully explored
in two, three and four dimensions and the complete equivalence
between them
was established. In the current correlators method, which we
developed here in detail, the bosonized
expressions are always given in terms of a bosonic
vector gauge field. These expressions
have the same structure in terms of the vacuum polarization
tensor in all dimensions considered here.
This remarkable property suggests that this method can be
extended for other higher dimensions. The corresponding bosonized
expressions are expected to have the same structure as found in our work,
at least in the weak coupling limit.
In the Master Lagrangean approach, on the other hand, bosonization
is made in terms a scalar field, vector gauge field and second
rank antisymmetric tensor gauge field, respectively, in two, three and
four dimensions. For arbitrary $n$-dimensions, it is therefore expected
that bosonization will be made in terms of an ($n-2$)-rank antisymmetric
tensor gauge field within this approach. Another possibility would be to
discuss the bosonization of other fermionic models like QED and the
Fermi theory of weak interactions. We intend to pursue these aspects
in a future work.

It is rather interesting to note that the topological structure of the
bosonized current found in the Master Lagrangean approach is the
only result which is always exact and depends neither on the interaction
nor on the dimensionality. Other operators, in dimensions higher than
two, are bosonized differently according to the dimension and the
interacting part of the theory they are embedded. This was explicitly
shown both for the current and the
kinetic part of the fermion lagrangean in the current correlators
approach. This is an important difference from the two-dimensional
bosonization, were expressions valid in the free case are also valid
in the presence of interaction. In fact this is the property which
allows for exact solvability of interacting two-dimensional models
through bosonization.

It is instructive to note that the present analysis enlightens and
unifies previous bosonization methods in higher dimensions.
The bosonization of the free massive fermion in
three dimensions considered in \cite{bfo}, for instance, corresponds to the
$B_\mu$-field as can be inferred from a comparison of the respective
lagrangeans.
Similarly, in the path integral bosonization
approach \cite{rb,rb1,bos}, the bosonic
field corresponds to  the one that appears in our Master Lagrangean analysis,
which is transparent by a comparison of the currents. It is interesting to
point out that in the infinite fermion mass limit in three dimensions,
the expressions for the bosonized lagrangean either in terms of the
$A_\mu$-field or the $B_\mu$-field are identical.
This is also the
only limit where such expressions are local \cite{rb,rb1,bos}.

Also in the case of the
free massless fermion bosonization in three dimensions,
for instance, performed in
\cite{em1} using a direct operator bosonization of the fermion field,
one can conclude by comparing
the structure of the
bosonic lagrangean and current found there, with the ones obtained in the
current correlator approach,
that the $B_\mu$-field is being used.
Furthermore, in the light of our analysis, we can easily understand why
the bosonization
formulas presented in \cite{bfo} interpolate between
the zero fermion mass \cite{em1} and the
infinite fermion mass \cite{rb, rb1,bos}
in the free theory in three dimensions,
even though the $B_\mu$-field is used in one case and the $A_\mu$-field
in the other. This is possible because, as remarked above, the 
methods based on the $A_\mu$-field and $B_\mu$-field coincide in the
infinite mass limit.

The observation that the operator bosonization introduced in \cite{em1}
is given in terms of the $B_\mu$-field, together with the structural
property found in the current correlator approach,
strongly suggests the possibility
of obtaining a direct Mandelstam-like operator bosonization of the fermion
field for the MTM in three and four dimensions for an arbitrary mass,
at least in the weak coupling limit, in terms of this field. This would
provide a complete bosonization scheme which is still lacking
in dimensions higher than two.

\bigskip
\bigskip

\leftline{\Large\bf Acknowledgements} \bigskip

Both authors  
were partially supported by CNPq-Brazilian National Research Council.
RB is very grateful to the Instituto de F\'\i sica-UFRJ 
for the kind hospitality.

\vfill
\eject

\end{document}